\documentclass[aps,twocolumn,showpacs,preprintnumbers,prl,amsmath,amssymb,amsfonts,superscriptaddress,floatfi
x]{revtex4-1}

\usepackage{caption}
\usepackage{amssymb,graphicx}
\usepackage[colorlinks=true, pdfstartview=FitV,
linkcolor=blue, citecolor=blue, urlcolor=blue]{hyperref}
\usepackage[utf8]{inputenc}
\usepackage{color}
\usepackage{lmodern}
\usepackage{float}
\usepackage[normalem]{ulem}
\usepackage{cancel}
\newcommand{\be}{\begin{equation}}
\newcommand{\ee}{\end{equation}}
\newcommand{\bea}{\begin{eqnarray}}
\newcommand{\eea}{\end{eqnarray}}

\begin{document}

\title{{Vortex supersolid in the XY model with tunable vortex fugacity}}
\author{I. Maccari}
\affiliation{Department of Physics, Stockholm University, Stockholm
  SE-10691, Sweden}
\author{N. Defenu}
\affiliation{Institute for Theoretical Physics, ETH Zürich,
  Wolfgang-Pauli-Str. 27, 8093 Zürich, Switzerland}
\author{C. Castellani}
\affiliation{Department of Physics, Sapienza University of Rome, P.le A. Moro 2, 00185 Rome, Italy}
\affiliation{Institute for Complex Systems (ISC-CNR), UOS Sapienza,
  P.le A. Moro 5, 00185 Rome, Italy}
\author{T. Enss}
\affiliation{Institute for Theoretical Physics, University of Heidelberg, 69120 Heidelberg, Germany}
\begin{abstract}
In this paper, we investigate the XY model in the presence of an additional potential term that independently tunes the vortex fugacity favouring their nucleation. By increasing the strength of this term and thereby the vortex chemical potential $\mu$, we observe significant changes in the phase diagram with the emergence of a normal vortex-antivortex lattice as well as a superconducting vortex-antivortex crystal ({lattice} supersolid) phase. We examine the transition lines between these two phases and the conventional non-crystalline one as a function of both the temperature and the chemical potential. Our findings suggest the possibility of a peculiar tricritical point where  second-order, first-order, and infinite-order transition lines meet. We discuss the differences between the present phase diagram and previous results for two-dimensional Coulomb gas models. Our study provides important insights into the behaviour of the modified XY model and opens up new possibilities for investigating the underlying physics of unconventional phase transitions.
\end{abstract}
\maketitle

\onecolumngrid
\section{Introduction}

Systems displaying multiple forms of long-range order in their ground state have always fascinated physicists for their potential to exhibit a complex phase diagram. Different from simpler systems, they can host multiple phase transitions and reveal new intermediate phases between the ground state and the high-temperature phase.
Apart from multi-component systems, such as multiband superconductors or bosonic mixtures, also single-component systems can present a similar scenario. 
The two-dimensional (2D) Coulomb gas (CG) model is a paradigmatic example. 

The 2D CG is an effective model for superconducting (SC) and superfluid vortices which, in two dimensions, are equivalent to logarithmically interacting charges. 
In the limit of small vortex fugacity, the model undergoes a Berezinskii-Kosterlitz-Thouless (BKT)~\cite{berezinskyDestructionLongrangeOrder1972,kosterlitzOrderingMetastabilityPhase1973, kosterlitzCriticalPropertiesTwodimensional1974} transition separating a low-temperature phase, where vortices and antivortices are tightly bound in pairs, from a high-temperature phase where free vortices proliferate and lead to a discontinuous vanishing of the condensate phase rigidity.

As the vortex fugacity $g$ increases above a critical value $g_c$, however, the low-temperature phase of the system undergoes a first-order phase transition from a vortex-vacuum superfluid to a vortex-antivortex superfluid crystal, which additionally breaks the discrete $Z_2$ symmetry associated with the two energetically equivalent checkerboard configurations of the lattice. 
As a result, in this regime the ground state exhibits two coexisting orders: a quasi-long-range order of the superfluid phase, characterized by a finite superfluid stiffness $J_s$, and a long-range positional order, characterized by a finite Ising order parameter for the staggered vorticity $M_{stag}$. 
Establishing how such a vortex supersolid melts into the disordered high-temperature phase has been a topic of great interest. 
The phase diagram of the 2D Coulomb gas at large vortex fugacity has been extensively investigated both for discrete lattice models ~\cite{leeNewCriticalBehavior1990, leeDenseTwodimensionalClassical1991, leePhaseTransitionsClassical1992} and in the continuum limit~\cite{lidmarMonteCarloSimulation1997}. In the presence of a discrete underlying grid, it was shown~\cite{leeNewCriticalBehavior1990, leeDenseTwodimensionalClassical1991, leePhaseTransitionsClassical1992} that at large vortex fugacity, the system undergoes two distinct phase transitions with an intermediate non-superfluid phase where the discrete $Z_2$ symmetry is spontaneously broken.

Addressing this problem within a 2D XY model has proven to be much more challenging. 
A ground state formed by a $Z_2$ vortex supersolid can be realized, in this model, by applying a uniform transverse magnetic field to the system with half a magnetic flux quantum crossing each plaquette of the spin lattice. The resulting model is the well-known fully frustrated XY (FFXY) model.  Over the years this has been the subject of extended theoretical discussions, with a series of conflicting analytical and numerical results about the number of phase transitions and their nature~\cite{teitelTwoDimensionalFullyFrustrated2013}. Finally, in 1996 Olsson~\cite{olssonTwoPhaseTransitions1995} numerically demonstrated the presence of two phase transitions that are very close together, with the BKT critical temperature, $T_{BKT}$, slightly smaller than the Ising critical temperature, $T_I$, associated with the vanishing of $M_{stag}$.
The theoretical argument for the observed splitting was afterwards provided by Korshunov~\cite{korshunovKinkPairsUnbinding2002}. The continuous nature of the Ising transition ensures that, when approaching $T_I$ from below, the proliferation of Ising domain walls with a net polarization continuously decreases both $M_{stag}$ and $J_s$. Hence, there are in general two possible scenarios that describe the melting of a ground state with coexisting superfluidity and staggered vortex structures: 1) the system exhibits a preemptive first-order phase transition with $J_s$ and  $M_{stag}$ vanishing discontinuously at the same critical temperature; 2) the system undergoes two phase transitions with $T_{BKT}<T_I$. Indeed, as soon as domain-wall excitations reduce $J_s$ below the BKT critical value $J_s(T_{BKT})= 2T_{BKT}/\pi$, vortex-antivortex pairs unbind and $J_s$ drops discontinuously to zero. 
The FFXY model exhibits the second scenario, as confirmed also by more recent numerical studies~\cite{hasenbuschTwodimensionalXYModel2005, okumuraSpinchiralityDecouplingCritical2011}.
Yet, although the ground state of the FFXY model shares the same orders and symmetries as that of the 2D CG model at large vortex fugacity $g$, neither the FFXY nor the classical XY model allows for a systematic study of the phase diagram as a function of $g$.
The XY model is, indeed, a single-coupling model where the value of the vortex fugacity cannot be tuned independently but is rather fixed by the value of the spin-exchange coupling $J$. 

In the present work, we face this challenge by studying the phase diagram of the modified XY model that we introduced in a previous work~\cite{maccariInterplaySpinWaves2020}, where the vortex fugacity can be tuned independently and in a direct way without changing the relevant interactions at play~\cite{duranVortexLatticeTwodimensional2020}. By employing large-scale Monte Carlo simulations we assess the phase diagram of the model and show that the system undergoes a single first-order phase transition with $T_{BKT}=T_I$ for a finite range of values of the vortex fugacity $g_c<g<g^*$, while for $g>g^*$ the two phase transitions split apart with $T_{BKT}<T_I$. The quantitative numerical characterisation of a BKT transition at large but finite vortex fugacity, which goes beyond the traditional BKT picture with a line of fixed points at zero fugacity, is relevant in numerous physical systems, 
including two-dimensional Kondo lattices~\cite{mizukamiExtremelyStrongcouplingSuperconductivity2011, Balatsky_heavyfermion2012}, and recently in the description of the metal-insulator transition in disordered 2D materials \cite{karcher2023}.  In thin superconducting films, a finite density of vortex-antivortex pairs can be induced at low temperatures by spatially correlated-disorder~\cite{maccariBroadeningBerezinskiiKosterlitzThoulessTransition2017, ilariamaccariBKTUniversalityClass2018}, while stable configurations of vortex supersolids can be realized via magnetic pinning arrays~\cite{milosevicVortexAntivortexLatticesSuperconducting2004, milosevicVortexantivortexLatticesSuperconducting2005} or superconductor/ferromagnet hybrid structures~\cite{bobbaVortexantivortexCoexistenceNbbased2014}. The formation and melting of a vortex-antivortex lattice in superfluid $^4$He films can be observed by the presence of a transverse mode that can exist only in the crystalline phase, and the vortex fugacity can be tuned by additional $^3$He atoms \cite{zhang1993}. More recent realisations include ultracold fermionic gases \cite{botelho2006} and polariton fluids \cite{hivet2014}.
High vortex fugacities may also emerge in long-range interacting systems. Indeed, generic power-law couplings $1/r^{\alpha}$ may disrupt the BKT in $d=2$ by increasing the vortex fugacity\,\cite{giachetti2021berezinskii,giachetti2022berezinskii}. It is worth noting that  $1/r^{2}$ interactions induce  BKT scaling also in several $d=1$ models\,\cite{kosterlitz1976phase,cardy1981one}.

\section{The model}

The model studied in this work is a modified version of the original XY model with an extra potential term  added to tune the vortex fugacity independently from the ferromagnetic coupling $J$. 
The Hamiltonian of the modified XY model, introduced in our previous work~\cite{maccariInterplaySpinWaves2020},  reads:

\begin{equation}
H_{XY}^{{\mu}}= -J\sum_{i, \nu=\hat{x},\hat{y}} \cos(\theta_i- \theta_{i+\nu})- {\mu} \sum_{i} \big( I_{P_i} \big)^2,
\label{H_2}
\end{equation}
with $I_{P_i}$ the spin current circulating around a unit plaquette $P_i$ of area $a^2=1$,
\begin{equation}
I_{P_i}=\sin(\theta_i - \theta_{i+\hat{x}}) + \sin(\theta_{i+\hat{x}}- \theta_{i+\hat{x}+\hat{y}})+ \sin(\theta_{i+\hat{x}+\hat{y}} - \theta_{i+\hat{y}}) + \sin(\theta_{i+\hat{y}} - \theta_{i}).
\label{current}
\end{equation}
For $\mu=0$, Eq.\eqref{H_2} is the classical XY model, where the value of the vortex fugacity is fixed by the bare spin stiffness $J$. On the other hand, by considering nonzero values of $\mu$ one can independently tune $g$ to either favour for $\mu>0$, or disfavour for $\mu<0$, the vortex nucleation in the system. Thus, by increasing $\mu>0$, the value of the vortex-core energy $\mu_v\propto -\mu$ decreases and, in turn, the value of the vortex fugacity $g=2\pi e^{-\beta \mu_v}$ increases.

The energy-entropy balance for the proliferation of free vortices suggests that the BKT critical temperature decreases as the value of $\mu$ increases. At the same time, it is also apparent that there exists a critical value $\mu=\mu_c$ at which the ground state of the system undergoes a first-order phase transition from a superfluid with vanishing vortex density $\rho_v(T\to 0)\to 0$ (``vortex vacuum'') to a vortex-antivortex superfluid crystal with $\rho_v(T\to 0)\to 1$~\cite{leeDenseTwodimensionalClassical1991}. 

While in our previous work~\cite{maccariInterplaySpinWaves2020} we focused on the regime $\mu<\mu_c$ 
here we will investigate the phase diagram of the model \eqref{H_2} for $\mu>\mu_c$.
{In this regime, the ground state is a checkerboard configuration of vortices and antivortices, forming a squared lattice. Being it superfluid, the ground state is a  ``supersolid'', or more precisely a ``lattice supersolid'', as the presence of an underlying square grid reduces the translational symmetry broken by the crystal from a continuous to a discrete $Z_2$ symmetry.~\cite{boninsegni_Supersolids2012, leonardSupersolidFormationQuantum2017}.}
As a function of $\mu$, we will determine the value of the two critical temperatures: $T_{BKT}$, at which a superfluid quasi-condensate forms, and $T_{I}$, at which a charge-ordered state forms, that is described by a real $Z_2$ order parameter 
associated with the two possible staggered magnetizations of the vortex-antivortex lattice.
This systematic investigation will enable us to assess the phase diagram of the system and to establish, for each value of $\mu$, whether the system displays two separate phase transitions, or a single preemptive first-order phase transition where both the superfluid stiffness $J_s$ and the staggered magnetization $M_{stag}$ jump discontinuously to zero at the same critical temperature $T_{BKT}= T_{I}$. 

\section{Monte Carlo simulations}

We assess the phase diagram of the model \eqref{H_2} in the regime $\mu>\mu_c$ via large-scale Monte Carlo (MC) simulations. This allows us to properly account for the non-trivial interactions between the different topological phase excitations at play, which include vortices, Ising-like domain walls between the two possible values of $M_{stag}$, and kink-antikink excitations along the domain walls~\cite{olssonKinkantikinkUnbindingTransition2005}.

We studied the model \eqref{H_2} on a discrete square grid of spacing $a=1$ and size $N= L\times L$, for different values of the linear size $L$.
Details of our MC simulations can be found in the Supplementary Materials.

To assess the values of the $BKT$ critical temperature, we computed the superfluid stiffness $J^{\nu}_s$, which measures the response of the system to a phase twist $\Delta_{\nu}$ along a given direction $\nu$.  This can be thought of in terms of twisted boundary conditions, $\theta_{i + L\hat{\nu} }= \theta_i + \Delta_{\nu}$, reabsorbed via a gauge transformation in a new set of variables $\theta_i^{'} = \theta_i - r_{i, \nu} \Delta_{\nu}/L$, with periodic boundary conditions.   For a superconducting film, it corresponds to the response to a transverse gauge field $\bf{A}$ and it signals the onset of perfect diamagnetism, i.e., the well-known Meissner effect. 
$J_s$ is defined as:
\begin{equation}
 J_{s}^{\nu} \equiv -\frac{1}{L^2}\frac{\partial^2 F(\Delta_{\nu})}{\partial A_{\nu}^2} \Big|_{A_{\nu}=0}
 \label{Js_def}
\end{equation}
and has two contributions
  \begin{equation}
    \label{eq:Jsx}
    J_s^{\nu} = J_d^{\nu} - J_p^{\nu},
  \end{equation}
 the diamagnetic ($J_d^{\nu}$) and the paramagnetic ($J_p^{\nu}$) response functions
\begin{align}
\label{Jd}
J_d^{\nu} &= \frac{1}{L^2} \Big[ \Bigl\langle \frac{\partial^2 H}{\partial A_{\nu}^2}\Big|_{0}\Bigr\rangle  \Big], \\
\label{jp}
J_{p}^{\nu} &= \frac{\beta}{L^2} \Big[ \Bigl\langle \Bigl(\frac{\partial H}{\partial A_{\nu}}\Big|_{0} \Bigr)^2\Bigr\rangle -\Bigl\langle \frac{\partial H}{\partial A_{\nu}}\Big|_{0}  \Bigr\rangle ^2 \Big],
\end{align}
where $\langle \dots \rangle$ stands for the thermal average over the MC steps. 
The explicit expressions of $J_d^{\nu}$ and $J_p^{\nu}$ are reported in the Appendix of~\cite{maccariInterplaySpinWaves2020}.
In this work, we have computed the superfluid response along $\nu \equiv \hat{x}$ and in what follows we will simply refer to $J_s \equiv J_s^{\hat{x}}$.

When increasing the temperature below $T_{BKT}$, the superfluid stiffness continuously decreases mainly due to the presence of non-topological phase excitations, such as spin waves and domain walls with a net polarization~\cite{korshunovKinkPairsUnbinding2002}. As soon as $T_{BKT}$ is reached, the proliferation of free vortices becomes entropically favoured and $J_s$ discontinuously jumps to zero.
According to the Nelson-Kosterlitz criterion~\cite{nelsonUniversalJumpSuperfluid1977}, at the critical point $J_s$ and $T_{BKT}$ are linked via the universal relation: $J_s(T_{BKT})=2 T_{BKT}/\pi$, which ultimately allows for the determination of the critical temperature. 

\begin{figure}[t!]
\centering
\includegraphics[width=\linewidth]{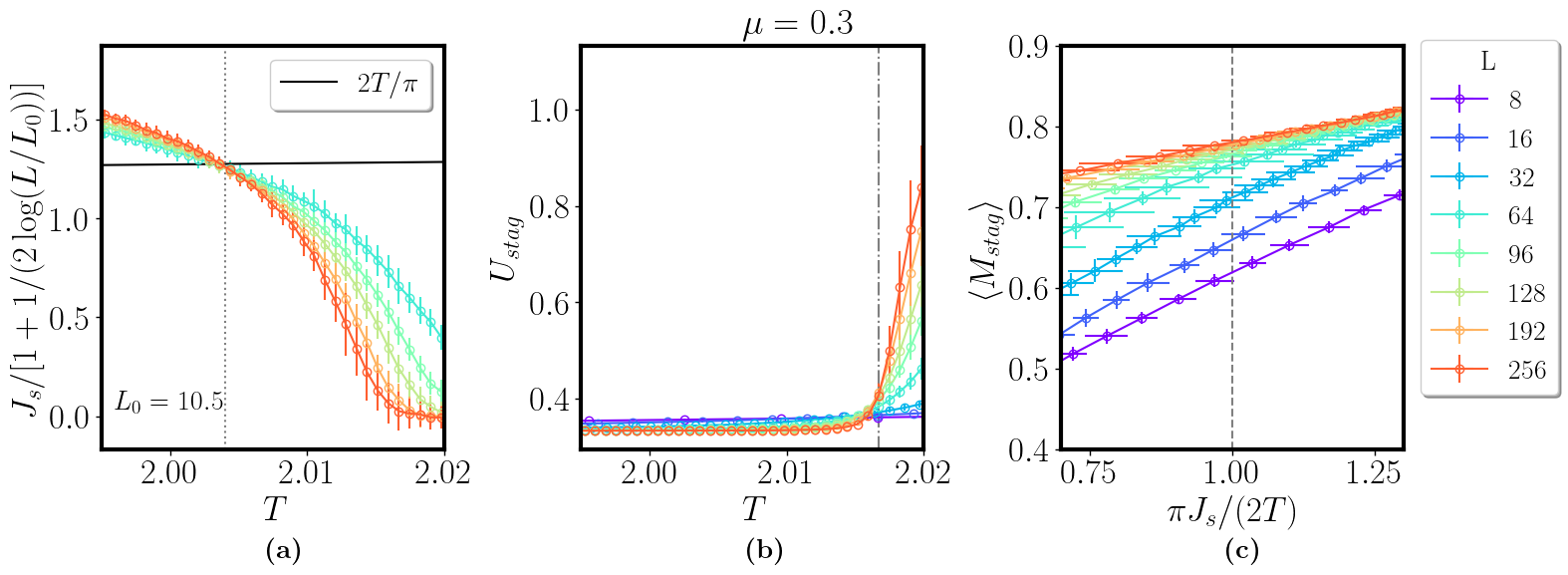}
\caption{Monte Carlo results for the case $\mu=0.3$. (a) Determination of $T_{BKT}$ from finite-size scaling of the superfluid stiffness $J_s$ renormalized according to the BKT scaling Eq.~\eqref{scaling_BKT} for $L=64, 96, 128, 192, 256$ (from top to bottom). The best crossing point is obtained with $L_0=10.5$. As expected, it lies on the $2T/\pi$ critical line (continuous black line). The dotted line indicates the extracted BKT critical temperature. (b) Determination of the Ising critical temperature $T_I$ from finite-size scaling of the Binder cumulant $U_{stag}$ defined in Eq.\eqref{ustagg}. The crossing point locates $T_I$, indicated here with a dashed-dotted line. (c) Olsson's plot~\cite{olssonTwoPhaseTransitions1995} for different values of the system size $L$. At the BKT critical point, while the superfluid stiffness jumps from $J_s(T_{BKT}^-)=2T_{BKT}^-/\pi$ to $J_s(T_{BKT}^+)=0$, the staggered magnetization is observed to increase with $L$ and reaches a finite value in the thermodynamic limit. This is an additional confirmation that $T_{BKT}<T_I$ in this case.  
The error bars are computed via a standard bootstrapping resampling method. Where not visible, the error bars are smaller than the point symbols.}
\label{mu0.3}
\end{figure}

In this work, we assess the value of $T_{BKT}$ by the BKT finite-size scaling of the superfluid stiffness~\cite{weberMonteCarloDetermination1988}:
\be
J_s(\infty, T_{BKT})=\frac{J_s(L, T_{BKT})}{1 + (2\log(L/L_0))^{-1}},
\label{scaling_BKT}
\ee
where $L_0$ is chosen to give the best crossing point at finite temperature (see also Supplementary Materials S2).  The BKT finite-size scaling of $J_s$ for $\mu=0.3 > \mu_c$ is reported in Fig.~\ref{mu0.3}(a), where we found $L_0=10.5$.

On the other hand, in order to assess the $Z_2$ Ising critical temperature $T_I$ associated with the melting of the vortex-antivortex crystal, we define a vortex ordering parameter as the \emph{staggered magnetization}:
\begin{equation}
 M_{stag} \equiv \sum_{i} (-1)^{x_i + y_i} q_i, 
 \label{mstagg}
\end{equation}
where $i$ labels the unitary plaquette of the spin lattice located at $(x_i, y_i)$. The vortex charge $q_i$ {is obtained by computing the phase circulation around each unitary plaquette, being:}
\begin{equation}
\hspace{-1cm}
    2\pi q_i= \left[ \theta_{i+\hat{x}} - \theta_i \right]^{\pi}_{-\pi} + \left[ \theta_{i+\hat{x} + \hat{y} } - \theta_{i+\hat{x}} \right]^{\pi}_{-\pi} 
 + \left[ \theta_{i+ \hat{y} } - \theta_{i+\hat{x} + \hat{y}} \right]^{\pi}_{-\pi} + \left[ \theta_{i} - \theta_{i + \hat{y} } \right]^{\pi}_{-\pi},
\end{equation}
{where the phase difference along each bond is defined so as to lie between the interval $\left(-\pi, \pi \right]$. The vortex charge}
 takes the values $q_i=0,+1,-1$, respectively, if a vortex, an antivortex, or zero vortices are located at the centre of the $i$-th plaquette. A vortex-antivortex crystal is characterised by $\langle M_{stag} \rangle = \pm 1$, according to the two possible equivalent configurations of the vortex-antivortex checkerboard.
To determine the value of $T_I$, we analyse the finite-size scaling of the Binder cumulant $U_{stag}$ associated with the staggered magnetization:
\begin{equation}
U_{stag}=\frac{\langle M_{stag}^4\rangle}{3 \langle M_{stag}^2\rangle ^2}.
 \label{ustagg}
\end{equation}
In the high-temperature limit the Binder cumulant approaches $U_{stag}(T \gg T_I) \to 1$ and in the low-temperature limit $U_{stag}(T \ll T_I) \to 0.3$, while at the critical point it is expected to assume a universal value independent on the system size~\cite{binderCriticalPropertiesMonte1981}. 
The finite-size scaling of the Binder cumulant is reported in Fig.\ref{mu0.3}(b) for $\mu=0.3$.

At this value of the vortex chemical potential $\mu=0.3$, we found two distinct and yet very close critical temperatures with $T_{BKT}=2.0040 \pm 0.0003 $ slightly smaller than  $T_I=2.01595\pm 0.00004$. 

As a further numerical confirmation of the splitting between the two phase transitions, we follow the scheme proposed by Olsson~\cite{olssonTwoPhaseTransitions1995}. Olsson's scheme consists in extracting a set of temperatures $T_{L}$ for different system sizes $L$, which are defined as the temperatures where the superfluid stiffness crosses the $2T/\pi$ BKT critical line, i.e., $J_s(T_L, L)=2T_L/\pi$. 
By increasing the size $L$, $T_L$ decreases and approaches the thermodynamic limit $T_{L\to \infty}\to T_{BKT}$ from above. If the two phase transitions are separated with $T_{BKT}<T_I$, the value of the staggered magnetization $\langle M_{stag}(T_L, L)\rangle$ at $T_L$ should increase with increasing system size $L$ and eventually reach a nonzero value in the thermodynamic limit. This is precisely what we observe in this case, as reported in Fig.~\ref{mu0.3}(c). At the temperatures $T_L$, indicated by a dashed vertical line, the value of  $\langle M_{stag}(T_L, L)\rangle$ increases, confirming that $T_{BKT}<T_I$.
To establish the phase diagram of the model \eqref{H_2}, we repeated the same analysis for different values of $\mu$. 

When approaching the critical value $\mu_c$ below which the ground state is a vortex-vacuum superfluid, we find that the separation between the two phase transitions reduces until they eventually merge into a single first-order phase transition at $\mu=\mu^*>\mu_c$. In particular, while down to $\mu=0.2$ (see Figs.~S2-S5 of the Supplementary Materials) we still find evidence of a splitting between the two transitions, at $\mu=0.175$ our numerical simulations suggest that the system undergoes a single first-order phase transition. 

The numerical evidence for a single first-order transition is threefold. The first indications in this sense are the failure of the BKT scaling Eq.~\eqref{scaling_BKT} for the superfluid stiffness (see Fig.~S6(a)) and the pronounced peaks in the Binder cumulant in the proximity of the critical point (see Fig.~S6(b))~\cite{vollmayrFiniteSizeEffects1993}. 

\begin{figure}[h!]
\centering
\includegraphics[width=\linewidth]{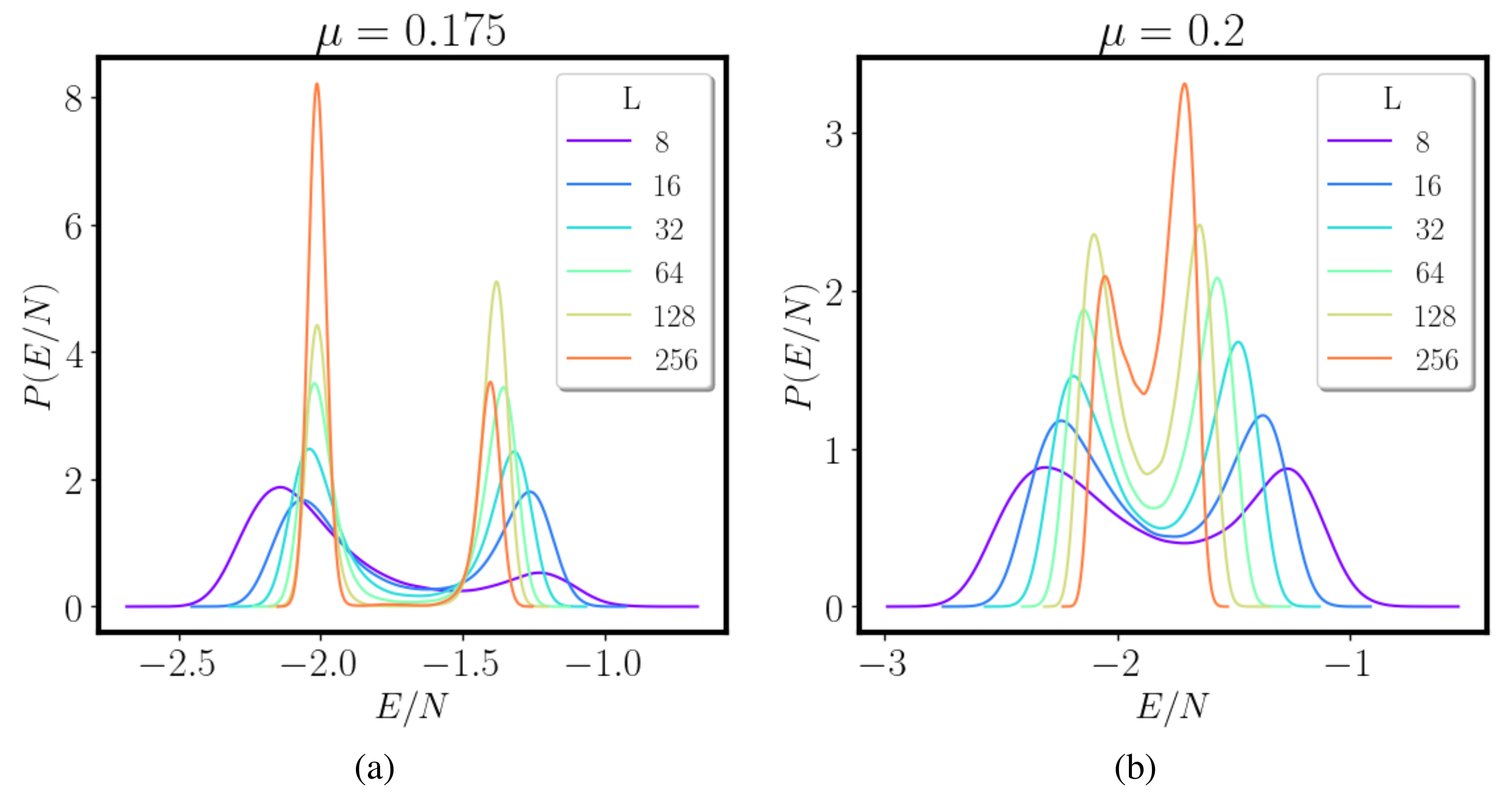}
\caption{Evidence for a first-order transition: the energy-density distribution $P(E/N)$ is shown for different system sizes $L$ ($N=L^2$) at the temperature corresponding to the specific-heat peak. While for (a) $\mu=0.175$ there are two peaks indicating a first-order transition, for (b) $\mu=0.2$ in the thermodynamic limit a single peak emerges consistent with a continuous transition second-order transition.}
\label{PE}
\end{figure}

 Second, an unambiguous demonstration of first-order phase transition at $\mu=0.175$ is provided by the presence of two peaks in the energy-density distribution $P(E/N)$ at the critical point. As reported in Fig.~\ref{PE}, at $\mu=0.175$ the minimum value $P(E_{min}/N)$ of the distribution between the two peaks vanishes by increasing the system size $L$ (see Fig.~\ref{PE}(a)), very differently from the case $\mu=0.2$ where $P(E_{min}/N)$ increases with $L$ (see Fig.~\ref{PE}(b)).

\begin{figure}[h!]
\centering
\includegraphics[width=\linewidth]{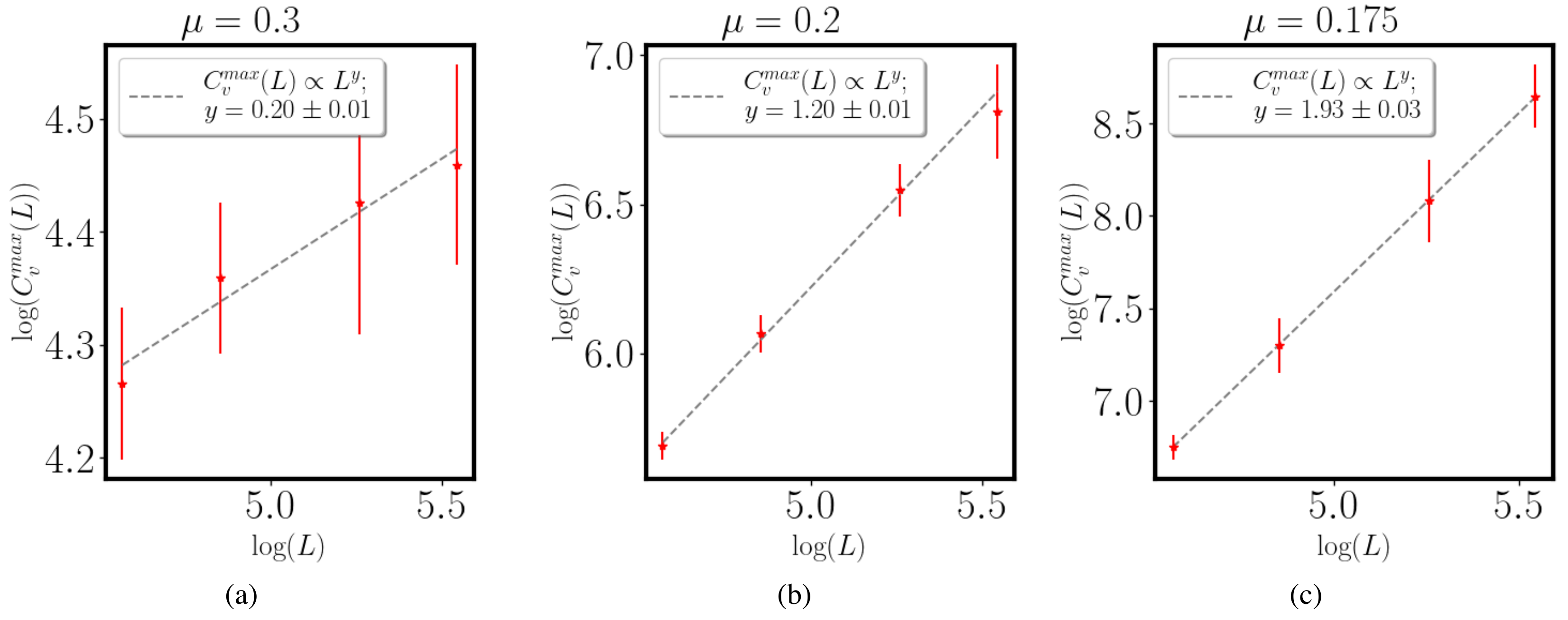}
\caption{Finite-size scaling analysis of the specific-heat peak $C_v^{max}$ at (a) $\mu=0.3$; (b) $\mu=0.2$; (c) $\mu=0.175$. The points in the three panels correspond to the linear sizes $L=96, 128, 192, 256$. }
\label{cvmax}
\end{figure}

Third, for a more quantitative analysis of the order of the transition, we looked at the finite-size scaling of the maximum value $C_v^{max}$ of the specific heat at the critical temperature. The specific heat $C_v$ being defined as:
\begin{equation}
    C_v= \frac{1}{T^2 L^2} \left( \langle E^2 \rangle - \langle E \rangle^2 \right),
\end{equation}
where $E$ is the total energy of the system.
For a second-order phase transition, $C_v^{max}$ scales as $C_v^{max} \propto L^{2/\nu -d}$, where $d=2$ is the spatial dimension of the system and $\nu=1$ is the critical exponent. 
Conversely, when the transition is of first order, for the Ising model in two dimensions the specific-heat peak scales as the volume of the system~\cite{vollmayrFiniteSizeEffects1993}, i.e., $C_v^{max} \propto L^{d}$. For $\mu=0.175, 0.2, 0.3$, we have extracted the value of $C_v^{max}$ at different system sizes $L$ and derived the exponent $C_v^{max} \propto L^{y}$ via a linear fit of the data in a log-log plot (see Fig.~\ref{cvmax}).
For $\mu=0.3$, this analysis yields $y=0.2\pm0.01$ (see Fig.~\ref{cvmax}(a)), in good agreement with the value $y=0$ 
expected in two spatial dimensions for a Ising-like second-order phase transition. For smaller $\mu$, instead, we observe a more divergent behaviour with $y=1.21 \pm 0.02$ at $\mu=0.2$ (see Fig.~\ref{cvmax}(b)) and, ultimately, $y=1.93\pm0.02$ for $\mu=0.175$, which is consistent with a first-order phase transition (see Fig.~\ref{cvmax}(c)).

Taken together, these findings consistently indicate the presence of a critical value $0.175\leq \mu^*<0.2$ at which the two phase transitions merge into a single first-order transition. At the same time, they also suggest the presence of a tricritical point $0.175\leq \mu_{\rm tric}<0.2$ at which the $Z_2$ second-order Ising transition becomes first order. Our data seem to indicate that for the modified XY model $\mu_{\rm tric}\equiv \mu^*$. At present, however, we cannot rule out the possibility that, although they are very close, $\mu_{\rm tric}>\mu^*$.

\begin{figure}[h!]
\centering
\includegraphics[width=\linewidth]{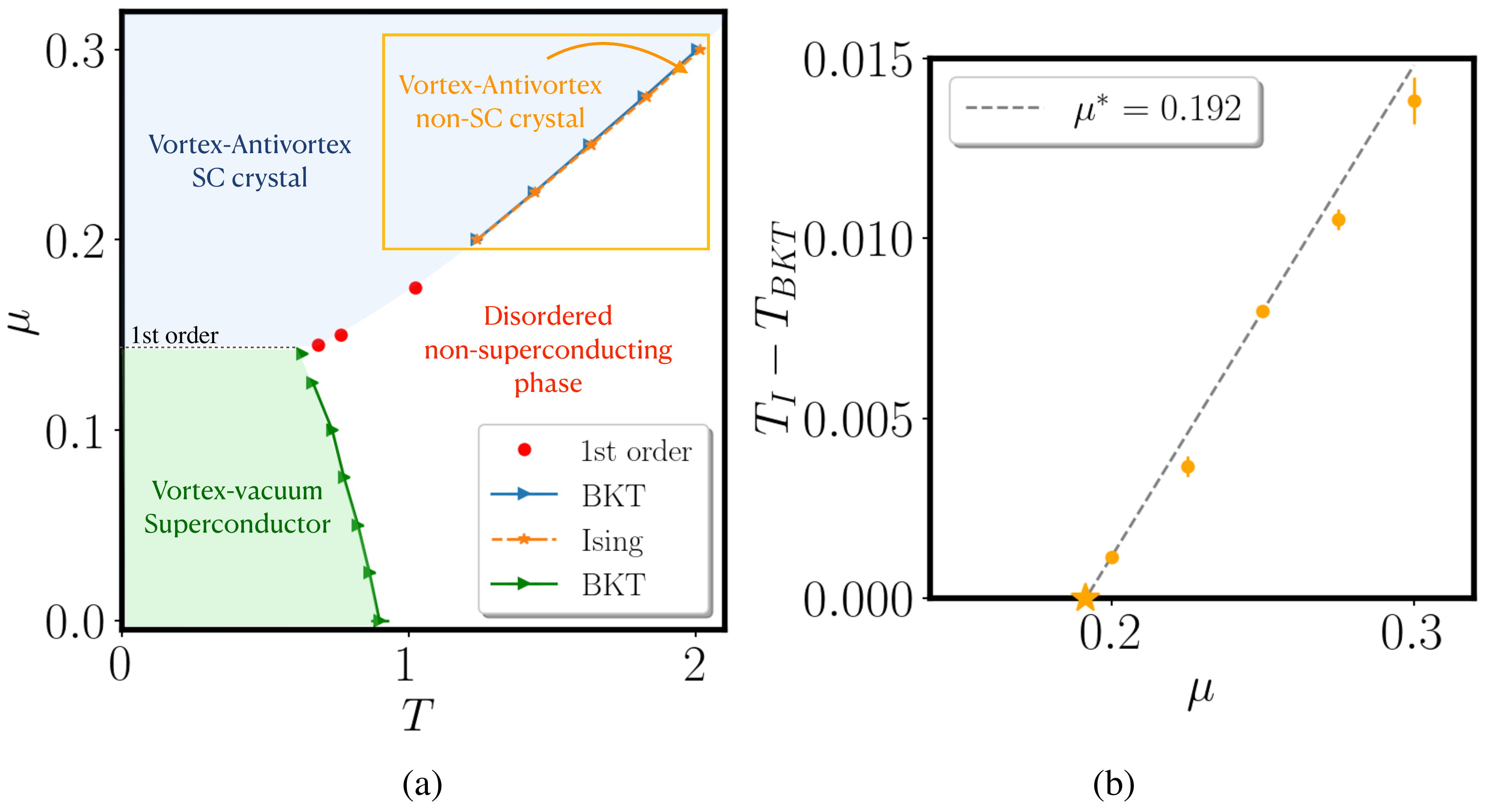}
\caption{(a) $\mu-T$ phase diagram of the model \eqref{H_2}. The light-blue and green areas indicate the two possible low-temperature states of the system. For $\mu \leq 0.14$ this is a vortex-vacuum superconducting state (green area), while for $\mu \geq 0.145$ it turns into a vortex-antivortex superconducting crystal (light-blue area). 
The BKT critical points (green triangles) in the region $\mu<\mu_c$ of the phase diagram are those derived in our previous work~\cite{maccariInterplaySpinWaves2020} and separate the vortex-vacuum SC state from the disordered high-temperature state. In the region $\mu_c<\mu<\mu^*$ the system undergoes a single first-order transition (red dots) from a vortex-antivortex SC crystal to a disordered non-SC state. Finally, for $\mu>\mu^*$ the two phase transitions separate and an intermediate non-SC state with a finite $Z_2$ crystalline order appears. To highlight the splitting of the two critical temperatures in this region of the phase diagram, in panel (b) we report the value of $T_I-T_{BKT}$ as a function of the chemical potential $\mu$. A linear fit of the data (dashed grey line) gives an estimate $\mu^*=0.192$ of the value at which the two phase transitions merge. 
The error bars are computed via a standard bootstrapping resampling method. Where not visible, the error bars are smaller than the point symbols.}
\label{phasediag}
\end{figure}

The complete phase diagram of the model \eqref{H_2} is shown in Fig.~\ref{phasediag}(a). For $\mu<\mu_c$ the BKT critical temperatures are those derived in our previous work~\cite{maccariInterplaySpinWaves2020}.
In the regime $\mu_c< \mu \leq 0.175$, the critical temperatures of the first-order phase transition have been computed by a finite-size scaling analysis of the temperatures corresponding to the specific-heat peak $C_v^{max}(T_c, L)$ (see Fig.~S7).

According to Fig.~\ref{phasediag}(a), for $\mu<\mu_c$ the system exhibits a single BKT phase transition from a quasi-long-range ordered superconducting state to a disordered one. By increasing the value of $\mu$ at low temperatures, the vortex fugacity increases until, at $0.14<\mu_c < 0.145$, the system undergoes a first-order phase transition~\cite{leeDenseTwodimensionalClassical1991} from a vortex-vacuum superconductor to a vortex supersolid which additionally breaks the $Z_2$ discrete symmetry associated with the two possible vortex-antivortex crystal configurations.

By increasing the chemical potential above the critical value $\mu_c$,  we find that up to a value of $\mu^*>\mu_c$ there exists a single first-order transition line separating the vortex-antivortex SC crystal from the high-temperature disordered state. For $\mu>\mu^*$, instead, the two phase transitions split apart with $T_{BKT} <T_I$. In this regime, a new intermediate phase appears where the system is a non-superconducting vortex-antivortex crystal spontaneously breaking the $Z_2$ symmetry associated with the charge ordering.  

Differently from the 2D Coulomb gas counterpart~\cite{leeNewCriticalBehavior1990}, however, the region of the phase diagram hosting this new phase is quite small and the two transitions remain close for all values of $\mu$ studied. Nonetheless, the splitting between the two transitions $\Delta T_c= T_I - T_{BKT}$  increases almost linearly with $\mu$ (see Fig.~\ref{phasediag}(b)). Via a linear fit of $\Delta T_c$ vs $\mu$, we also extracted an estimate of $\mu^*$ at which the two transitions merge. The obtained value  $\mu^*=0.192$ is consistent with the analysis reported above. 

\section{Conclusions}

In this study, we conducted a comprehensive numerical investigation of the modified XY model by introducing a plaquette term to control the fugacity of vortices. Our findings reveal that as the vortex fugacity increases, the low-temperature superfluid BKT state turns into a vortex supersolid
with finite superconducting density and charge ordering. 
At low temperatures, this state emerges from the superconducting vacuum via a first-order phase transition. However, as the temperature increases, a complex phase diagram emerges. At temperatures $T\lesssim 1$ and chemical potential $\mu\leq0.14$, a BKT transition line branches out of the first-order line, and vortex unbinding destroys the superconducting order. The transition line separating this new disordered state from the superconducting crystal remains first order up to $\mu^{*}\approx \mu_{\rm tric}$, while for larger $\mu$ an increasing temperature leads to the vanishing of superfluid order via the BKT mechanism, followed by the melting of the normal vortex-antivortex crystal into the disordered state via an Ising-like second-order line, as shown in Fig.\,\ref{phasediag}.

Our results are consistent with the analysis conducted in Ref.\,\cite{leeDenseTwodimensionalClassical1991} for the two-dimensional Coulomb gas, but two important differences stand out:
\begin{enumerate}
    \item First, the area between the two transition lines separating the superconducting crystal from the normal crystal and the disordered state is extremely small and only grows linearly by increasing the chemical potential.
    \item Second, the branching point of the second BKT line coincides within our numerical precision with the tricritical point $\mu_{\rm tric}$, where the first-order line meets the second-order Ising transition. 
    \end{enumerate}
    These differences may be attributed to the intrinsic differences between the two Hamiltonians, particularly to the fact that the topological excitations, i.e., the vortices, are coupled to the low-energy spin waves in the XY model, while this interaction is neglected in the Coulomb gas representation of the problem. Additionally, while our study focuses primarily on the superfluid stiffness $J_s$, Ref.\,\cite{leeDenseTwodimensionalClassical1991} characterizes the superconductor by the inverse dielectric constant. These two quantities are closely related in the traditional XY model with $\mu=0$, but the same relation does not hold in this study, where the plaquette term in the Hamiltonian \eqref{H_2} gives an explicit contribution to the superfluid stiffness.

In conclusion, resolving the nature of the unconventional tricritical point, where the first- and second-order lines meet with the infinite-order BKT line, requires the derivation of an improved BKT flow equation that can capture the mechanism of defect unbinding at finite fugacity. Such a theoretical framework should be able to capture both BKT scaling and the second-order transition line within the same formalism, and its development represents the most significant future direction of this work.
{At the same time, research on chiral magnets with strong Dzyaloshinskii-Moriya interactions (see \cite{duranVortexLatticeTwodimensional2020} and references therein), as well as experimental realizations of cold-atom systems with related phase diagrams~\cite{Loida_BondOrderWave2017} or spin-torque interactions~\cite{Ferraretto_Chiraledge2023}, can provide a complementary experimental route to investigate the nature of such unconventional tricritical point.
}

\section*{Acknowledgements}

The simulations were performed on resources provided by the Swedish National Infrastructure for Computing (SNIC) at the National Supercomputer Center at Link\"oping, Sweden. I.M. acknowledges the Carl Trygger foundation through grant number CTS 20:75. This work is supported by the Deutsche Forschungsgemeinschaft (DFG, German Research Foundation) under project-ID 273811115 (SFB1225 ISOQUANT) and under Germany’s Excellence Strategy EXC2181/1-390900948 (the Heidelberg STRUCTURES Excellence Cluster).

\bibliography{maccari}

%\clearpage
\onecolumngrid
\appendix
\section{Supplementary information}
\setcounter{figure}{0}
\renewcommand\thefigure{S\arabic{figure}}

\section{S1. Details of the Monte Carlo simulations}

In our simulations, a single MC step consists of the Metropolis sweeps of the whole lattice of spins. To let the system thermalize faster at low temperatures, we implemented a parallel tempering algorithm, allowing a swap of the spin configurations between neighbouring temperatures. Here, we propose one set of swaps after 32 MC steps. For each value of $\mu$ and $L$ simulated, we performed a total of $3 \times 10^5$ Monte Carlo steps, discarding the transient time occurring typically within the first $100000$ steps.

\clearpage

\section{S2. Assessing the Berezinskii-Kosterlitz-Thouless transition }

The BKT critical point can be located by finite-size scaling of the superfluid stiffness~\cite{weberMonteCarloDetermination1988}:

\begin{equation}
    \frac{\pi J_s(L, T_{BKT})}{2 T_{BKT}} = 1 + \frac{1}{ 2 \ln(L/L_0)}; 
    \label{BKT_scaling1}
\end{equation}

that can be rewritten as:
\begin{equation}
J_s(\infty, T_{BKT})=\frac{J_s(L, T_{BKT})}{1 + (2\log(L/L_0))^{-1}}.
%\label{scaling_BKT}
\end{equation}
 
\begin{figure}[ht!]
\centering
\includegraphics[width=\linewidth]{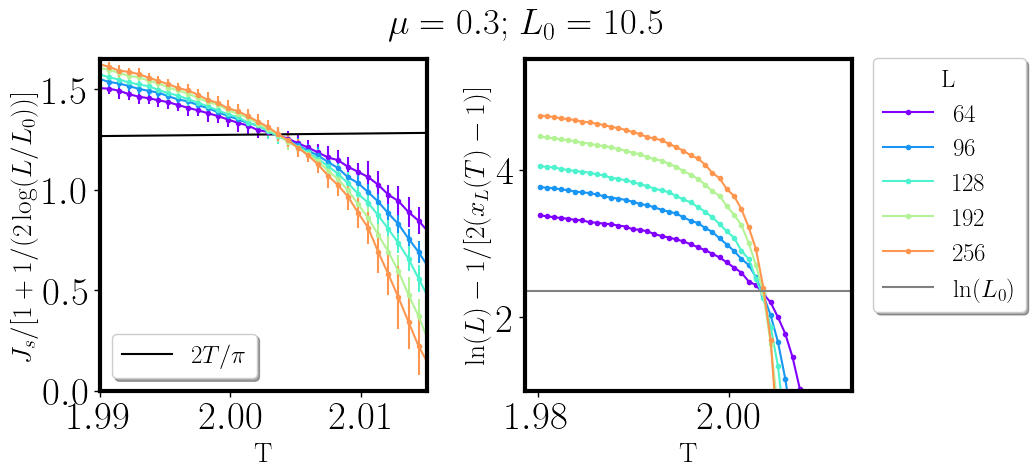}
\caption{(a) Superfluid stiffness $J_s$ rescaled according with the BKT scaling function of Eq.\eqref{scaling_BKT}. As expected, the crossing point lies on the BKT critical line $2T/\pi$. (b) Extrapolation of the parameter $L_0$ via the finite-size scale crossing of the function $f(L, T)$.}
\label{mu03L0}
\end{figure}

In the present analysis, we extrapolated $L_0$ using the BKT scaling itself so as to avoid undesired inconsistencies. Indeed, by rewriting Eq.\eqref{BKT_scaling1} as:

\begin{equation}
    \ln(L) - \frac{1}{2 (x_L(T_{BKT}) - 1)} = \ln(L_0),
    \label{BKT_scaling2}
\end{equation}

where $x_L(T) = \frac{\pi J_s(L, T)}{2 T}$, the crossing point of the function:
\begin{equation}
    f(L, T)= \ln(L) - \frac{1}{2(x_L(T) -1)},
    \label{L0}
\end{equation}
plotted as a function of the temperature for different values of $L$, can be used to directly extrapolate the value of $L_0$. 
We use this procedure to obtain the value of $L_0$, as shown for the case $\mu=0.3$ in Fig.\ref{mu03L0}. 

% The values of $L_0$ extracted as a function of $\mu>\mu^*$ are shown in Fig. S\ref{L0_mu}.

%\vspace{4cm}
\clearpage

\section{S3. Finite-size analysis for $\mu>\mu^*$ }

\begin{figure}[ht!]
\centering
\includegraphics[width=\linewidth]{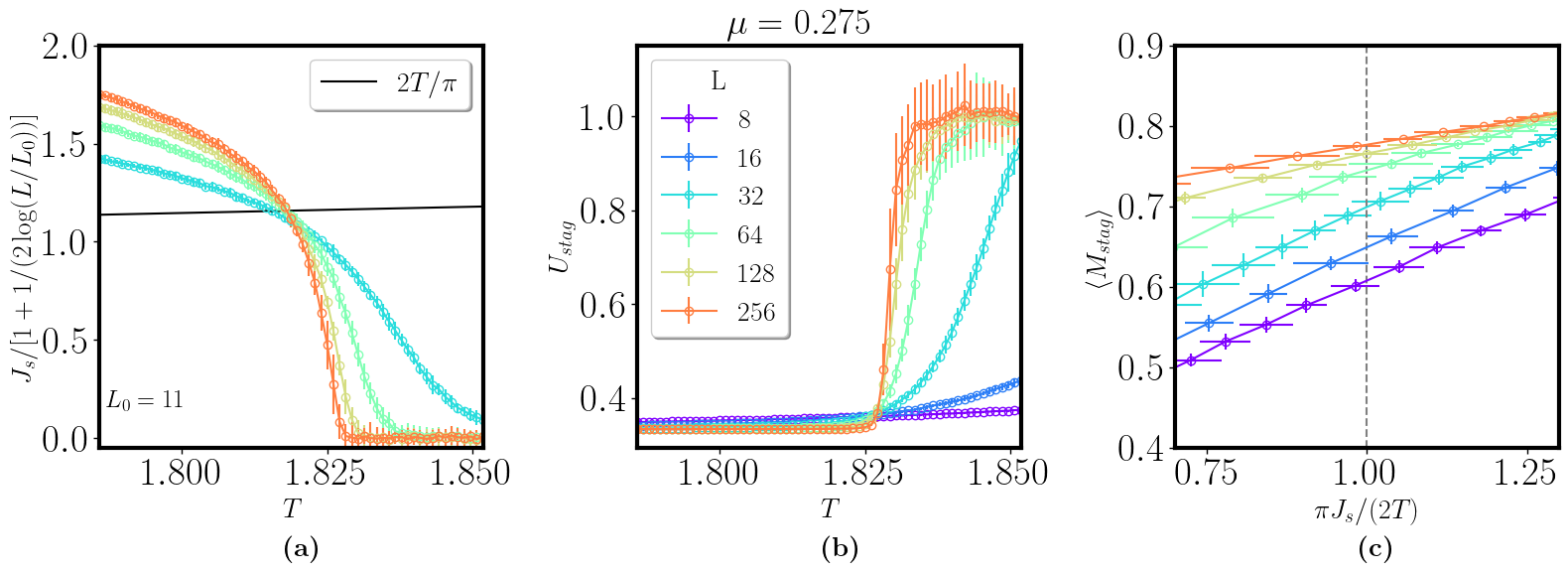}
\caption{Monte Carlo results for the case $\mu=0.275$. (a)Finite-size scaling of the superfluid stiffness $J_s$ renormalized according to the BKT scaling Eq.\eqref{scaling_BKT}. (b) Finite-size scaling of the Binder cumulant $U_{stag}$. (c) Finite-size Olsson's plot~\cite{olssonTwoPhaseTransitions1995}. It shows that at the BKT critical point, the staggered magnetization is finite in the thermodynamic limit. Thus confirming that $T_{BKT}<T_I$.}
\label{mu0.275}
\end{figure}

\begin{figure}[ht!]
\centering
\includegraphics[width= \linewidth]{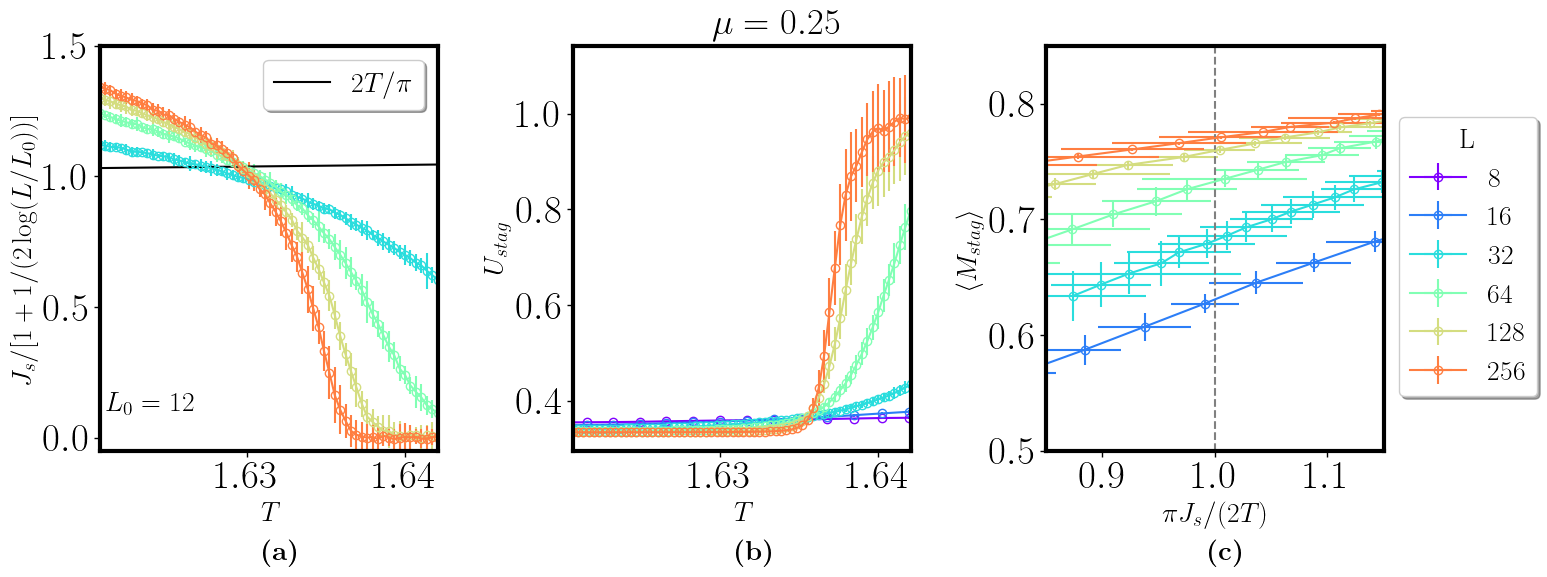}
\caption{Monte Carlo results for the case $\mu=0.25$. (a)Finite-size scaling of the superfluid stiffness $J_s$ renormalized according to the BKT scaling Eq.\eqref{scaling_BKT}. (b) Finite-size scaling of the Binder cumulant $U_{stag}$. (c) Finite-size Olsson's plot~\cite{olssonTwoPhaseTransitions1995}. It shows that at the BKT critical point, the staggered magnetization is finite in the thermodynamic limit. Thus confirming that $T_{BKT}<T_I$.}
\label{mu0.25}
\end{figure}

\begin{figure}[ht!]
\centering
\includegraphics[width=\linewidth]{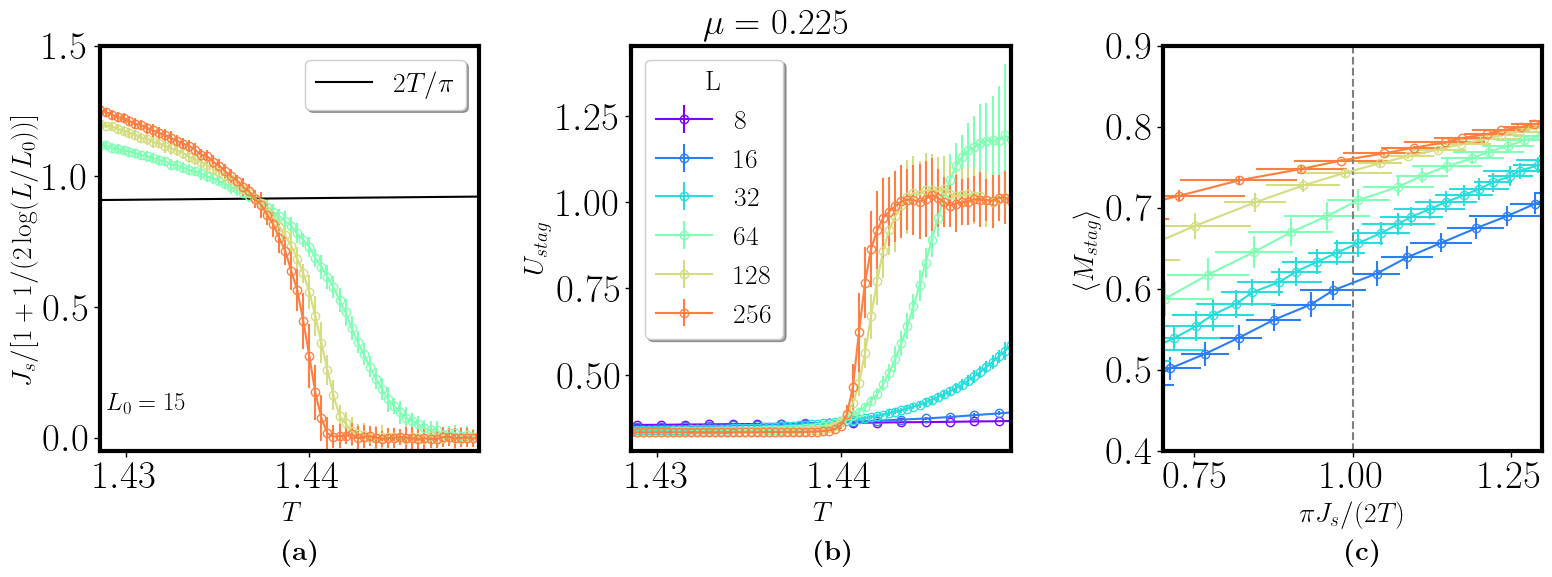}
\caption{Monte Carlo results for the case $\mu=0.225$. (a)Finite-size scaling of the superfluid stiffness $J_s$ renormalized according to the BKT scaling Eq.\eqref{scaling_BKT}. (b) Finite-size scaling of the Binder cumulant $U_{stag}$. (c) Finite-size Olsson's plot~\cite{olssonTwoPhaseTransitions1995}. It shows that at the BKT critical point, the staggered magnetization is finite in the thermodynamic limit. Thus confirming that $T_{BKT}<T_I$.}
\label{mu0.225}
\end{figure}

\begin{figure}[ht!]
\centering
\includegraphics[width=\linewidth]{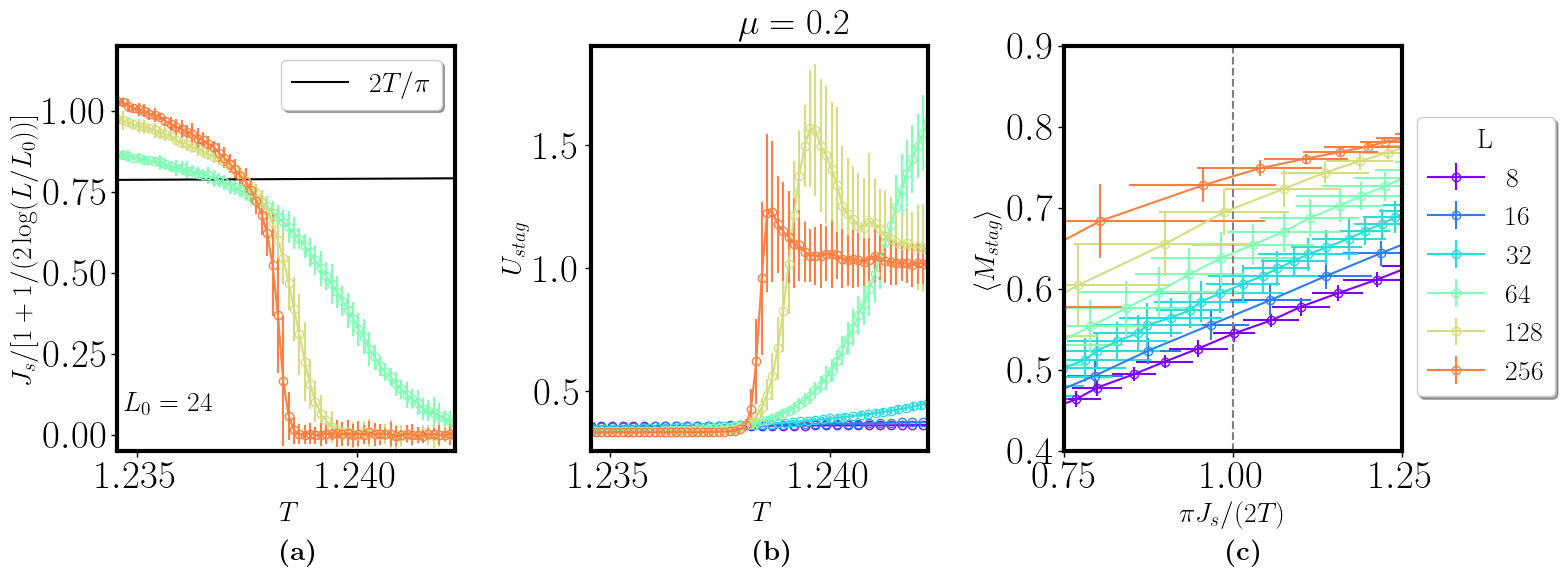}
\caption{Monte Carlo results for the case $\mu=0.2$. (a)Finite-size scaling of the superfluid stiffness $J_s$ renormalized according to the BKT scaling Eq.\eqref{scaling_BKT}. (b) Finite-size scaling of the Binder cumulant $U_{stag}$. (c) Finite-size Olsson's plot~\cite{olssonTwoPhaseTransitions1995}. It shows that at the BKT critical point, the staggered magnetization is finite in the thermodynamic limit. Thus confirming that $T_{BKT}<T_I$.}
\label{mu0.2}
\end{figure}

\clearpage

\section{S4. Preliminary indications of a first-order phase transition at $\mu=0.175$.}

\begin{figure}[ht!]
\centering
\includegraphics[width=\linewidth]{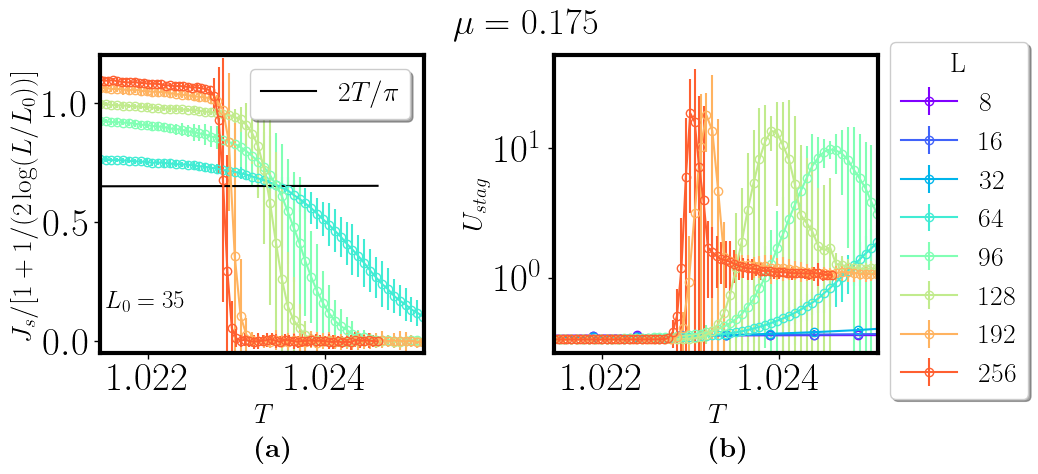}
\caption{The Monte Carlo results for the case $\mu=0.175$ show evidence of a single first-order phase transition. The preliminary indications have been: (a) the failure of the BKT scaling for the superfluid stiffness $J_s$; (b) a pronounced peak in the temperature dependence Binder cumulant. }
\label{mu0.175}
\end{figure}

\section{S5. Extrapolation of the critical temperatures in the regime $\mu_c<\mu<\mu^*$ }

\begin{figure}[ht!]
\centering
\includegraphics[width=\linewidth]{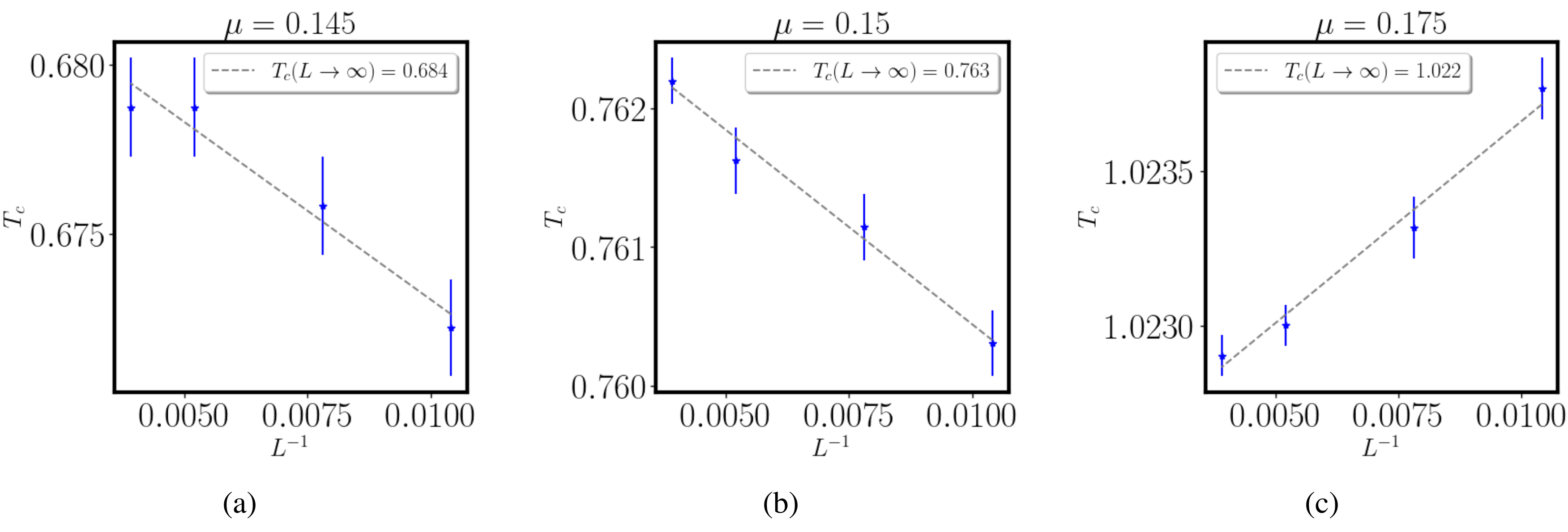}
\caption{Extrapolation of the critical temperature $T_c$ via a finite-size scaling analysis of the specific-heat peak. The points shown correspond to the linear system sizes $L= 96, 128, 192, 256$ respectively for (a) $\mu=0.145$; (b)$\mu=0.15$ (c) $\mu=0.175$. }
\label{Tc_Cv}
\end{figure}
\clearpage

\end{document}